\documentclass[useAMS,usenatbib]{mn2e}

\usepackage{graphicx}

\title[The nucleus of the Circinus galaxy]{Mid-infrared, spatially-resolved spectroscopy of the nucleus of the Circinus galaxy}
\author[P.F. Roche et al.]{Patrick F. Roche$^{1}$\thanks{E-mail:
p.roche@physics.ox.ac.uk },  Christopher Packham$^{2}$,  Charles M. Telesco$^{2}$,
\newauthor James T. Radomski$^{3}$, Almudena Alonso-Hererro$^{4}$, David K. Aitken$^5$, 
\newauthor  Luis Colina$^{4}$,  Eric Perlman$^{6}$, 
 \\
$^{1}$Astrophysics, University of Oxford, Dept of Physics, DWB, Keble Road Oxford OX1 3RH\\
$^{2}$Department of Astronomy, University of Florida, 211 Bryant Space Science Center, PO Box 112055, Gainesville, Fl 32611-2055, USA. \\
$^{3}$Gemini Observatory, c/o AURA, Casilla 603, La Serena, Chile. \\
$^{4}$Departamento de Astrof\'{\i}sica Molecular e Infrarroja, Instituto de Estructura de la Materia, Serrano 113b, Madrid, Spain. \\
$^{5}$ Dept.Physical Sciences,  University of Hertfordshire, College Lane, Hatfield, HERTS, AL10 9AB. \\
$^{6}$ Physics Department,  University of Maryland,  1000 Hilltop Circle, Baltimore, MD 21250 \\
}
\begin{document}

\date{Accepted  . Received 2005 December XX; in original form 2005 October 17}

\pagerange{\pageref{firstpage}--\pageref{lastpage}} \pubyear{2006}

\maketitle

\label{firstpage}

\begin{abstract}

High spatial resolution spectroscopy at 8-13$\umu$m with T-ReCS on Gemini-S has revealed striking variations in the mid-infrared emission and absorption in the nucleus of the Circinus galaxy on sub-arcsecond scales.  The  core of Circinus is compact and obscured by a substantial column of cool silicate dust.  Weak extended emission to the east and west coincides with the coronal line region and arises from featureless dust grains which are probably  heated by line emission in the coronal emission zone.  The extended emission on the east side of the nucleus displays a much deeper silicate absorption than that on the west, indicating significant columns of cool material along the line of sight and corresponding to an additional extinction of A$_{\rm V} \sim$25 mag. Emission bands from aromatic hydrocarbons are not subject to this additional extinction,  are relatively weak in the core and in the coronal line region, and are much more spatially extended than the continuum dust emission; they presumably arise in the circumnuclear star-forming regions. These data are interpreted in terms of an inclined disk-like structure around the nucleus extending over tens of parsecs and possibly related to the inner disk found from observations of water masers by Greenhill et al (2003). 

\end{abstract}


\section{Introduction}

The Circinus galaxy, hereafter Circinus, is a nearby spiral galaxy (distance $\sim$4~Mpc Freeman et al 1978) lying behind the Galactic plane and inclined at about 65 degrees to our line of sight.   It suffers significant Galactic interstellar extinction (A$_V \sim$ 1.7mag), but is seen clearly at infrared wavelengths.  The active nucleus in the core of the Circinus galaxy is one of the closest Seyfert galaxy nuclei to the Earth, allowing high spatial resolution investigations of the central activity and circumnuclear material; at a distance of 4~Mpc, 1 arcsec corresponds to 20~pc. 

The infrared spectrum of the nucleus is dominated by dust emission but shows a deep minimum near 10~$\umu$m (Moorwood and Glass, 1984). The mid-infrared spectrum measured in a small (4 arcsec diameter) aperture from the ground shows that the nuclear spectrum has a deep silicate absorption band (Roche et al 1991) indicating large columns of cool material along the line of sight. However, the spectrum measured in a large ($\sim$20 arcsec)  aperture with the Infrared Space Observatory displays prominent narrow emission bands attributed to aromatic hydrocarbon emission (Moorwood et al 1996).  The small  beam used in the ground-based observations isolates the emission from the nucleus while the large ISO beam includes significant contamination from the circumnuclear star forming regions.  The presence of the active nucleus is betrayed by strong infrared high excitation emission lines (Oliva et al 1994, Moorwood et al 1996, Maiolino et al 1998) indicating a much harder exciting spectrum than can be provided by star formation regions.  X-ray observations are consistent with a large absorbing column of material (n$_{\rm H}$  $\sim4 \times 10^{24} $cm$^{-2}$) towards the nucleus, (Matt, Brandt \& Fabian 1996), while the detection of broad H-alpha emission in polarised flux indicates an obscured broad line region (Oliva, Marconi \& Moorwood 1999).  All of these observations are consistent with the classification of Circinus as a Seyfert-II nucleus in which the broad line region is obscured from our direct  view by optically and geometrically thick circumnuclear material. 

\begin{figure}
   \centering
   \resizebox{\columnwidth}{!}{\includegraphics[clip=true]{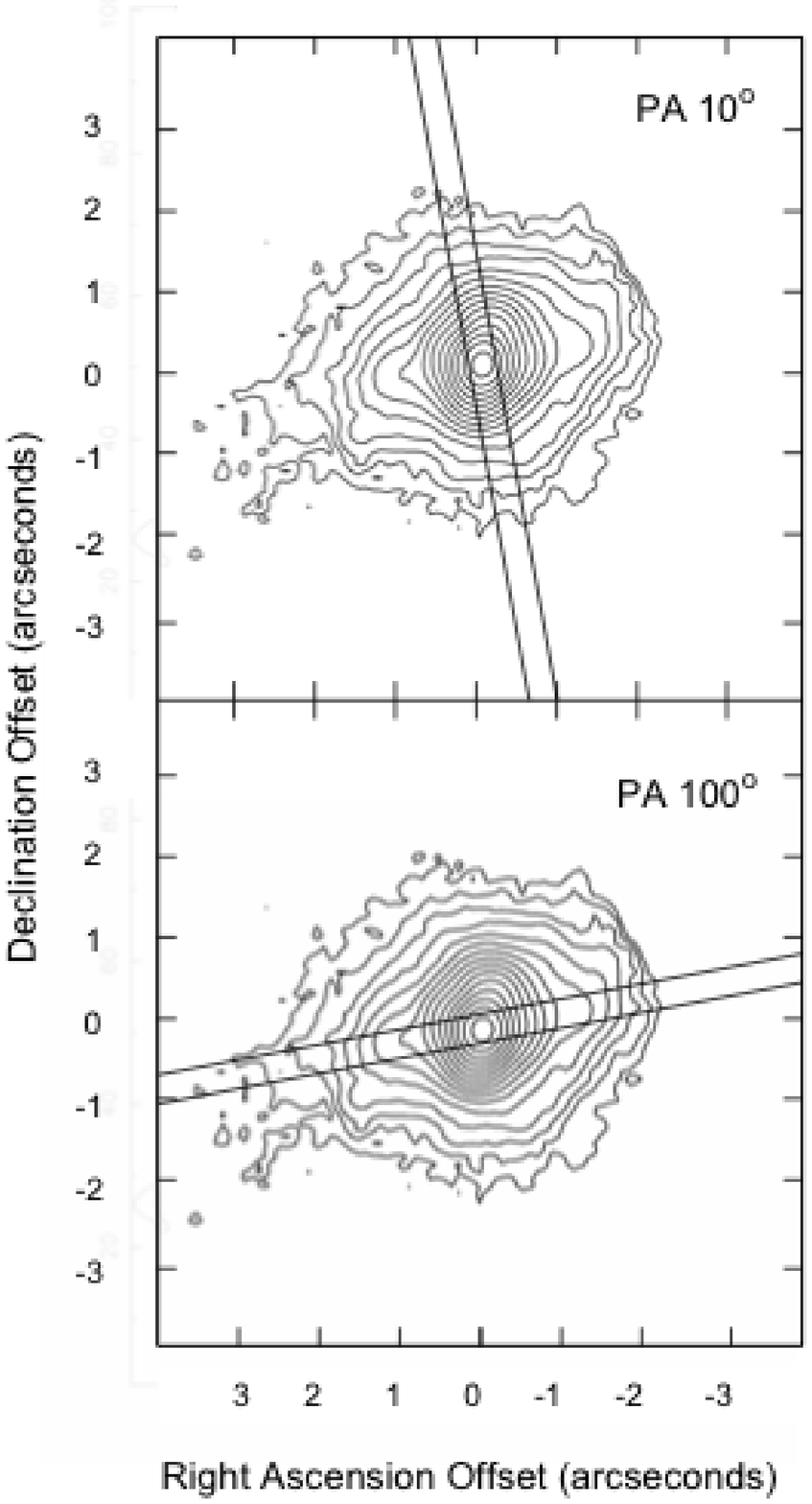}
   }
   \vspace*{-10mm}
     \caption{The T-ReCS slit positions superimposed on the 12$\umu$m Circinus acquisition image.   The image is centred at the peak of the mid-infrared emission, which is assumed to coincide with the active nucleus; the raw telescope coordinates give this position as RA = 14 13 09.3, Dec= -65 20 20 (J2000) . The contours are logarithmic and separated by a factor of 1.5.   North is up, East to the left. }  
        \label{slits}
    \end{figure}

\begin{figure*}
   \centering
   \resizebox{\hsize}{!}{\includegraphics[clip=true]{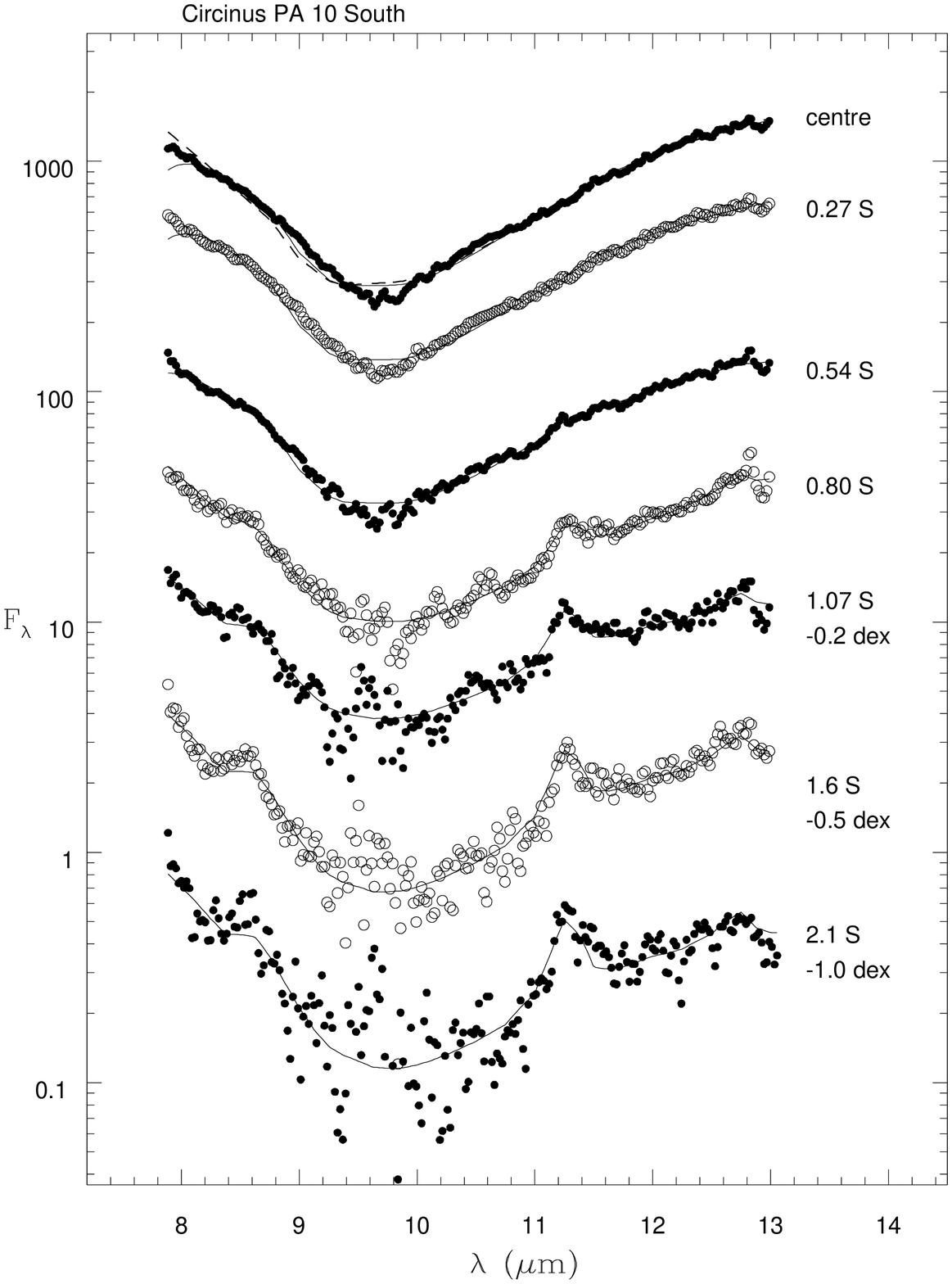}
 \includegraphics[clip=true]{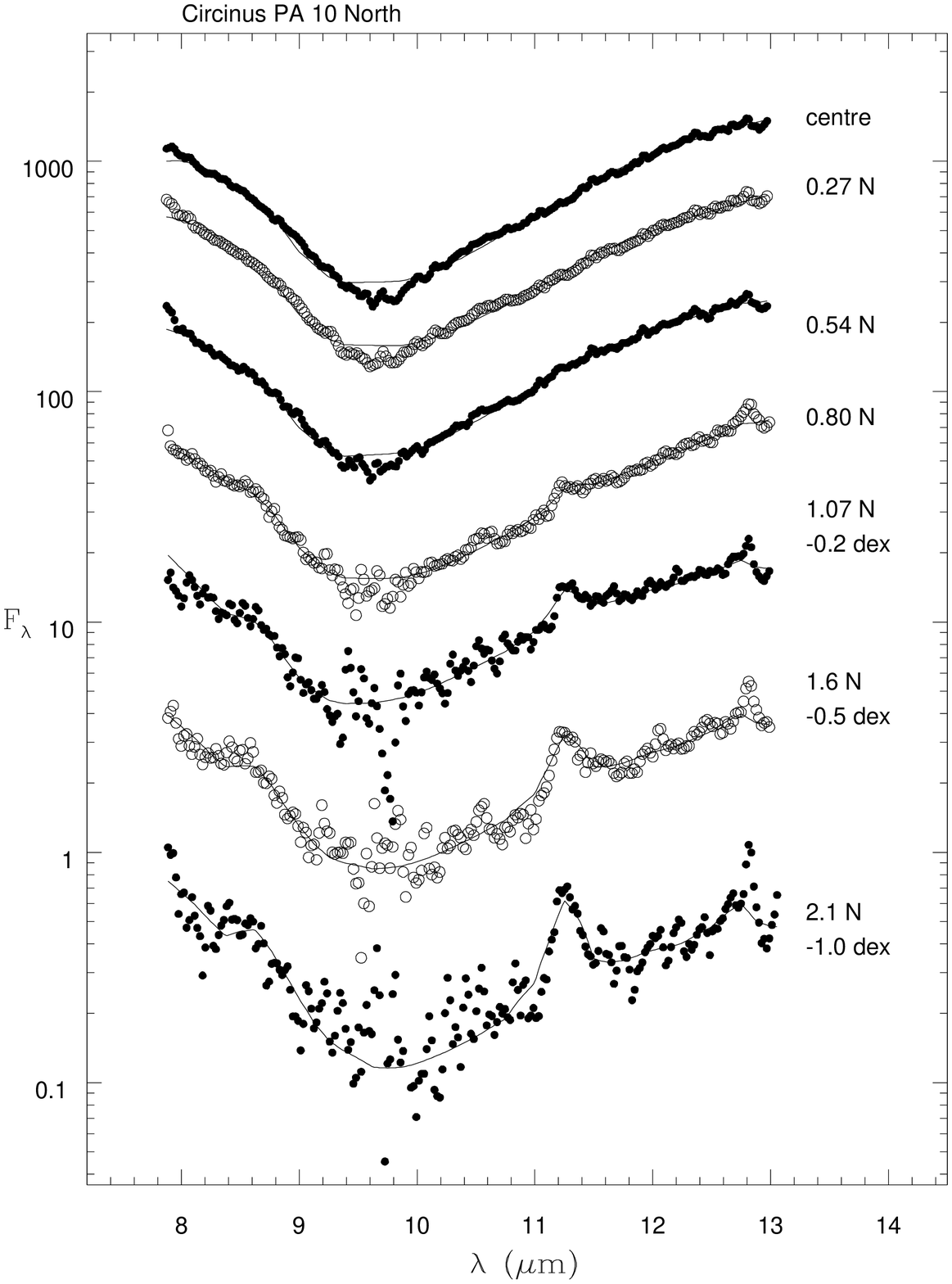}
}
     \caption{The spectra obtained at PA 10 degrees. The spectra on the south and north sides of the nucleus are shown at left and right respectively.  Note that there is a change in the spatial offset at 1.07 arcsec separation and that the on-nucleus spectrum is shown in both panels. The line drawn through the spectra represents the fits described in the text.  Flux is in units of 10$^{-20}$Wcm$^{-2} \umu$m$^{-1}$. 
}  
        \label{spec10}
    \end{figure*}

Kinematic and spatial information from observations of water masers over the central 0.4 pc indicate that there is a substantial amount of cool material ($\le$4 10$^5$ M$_\odot$) in approximately Keplerian orbit around a 1.7 10$^6$M$_\odot$ central mass (Greenhill et al 2003). This is interpreted as  a warped disk encircling the nucleus, approximately orthogonal to the ionisation cone detected in the visible (e.g. Wilson et al. 2000).

Recent 8 and 18~$\umu$m images have shown a strong point-like nuclear source accompanied by weaker extended emission oriented approximately east-west (Packham et al. 2005) and coincident with the coronal line emission region which is extended over ~2 arcsec and traces the high-excitation zone around the nucleus  (Maiolino et al 1998, 2000, Prieto et al 2004).  Here we present spatially resolved 8-13~$\umu$m spectra to investigate the nature of this extended emission and its relation to the active nucleus.

\section[]{Observations}

\begin{figure*}
   \centering
   \resizebox{\hsize}{!}{\includegraphics[clip=true]{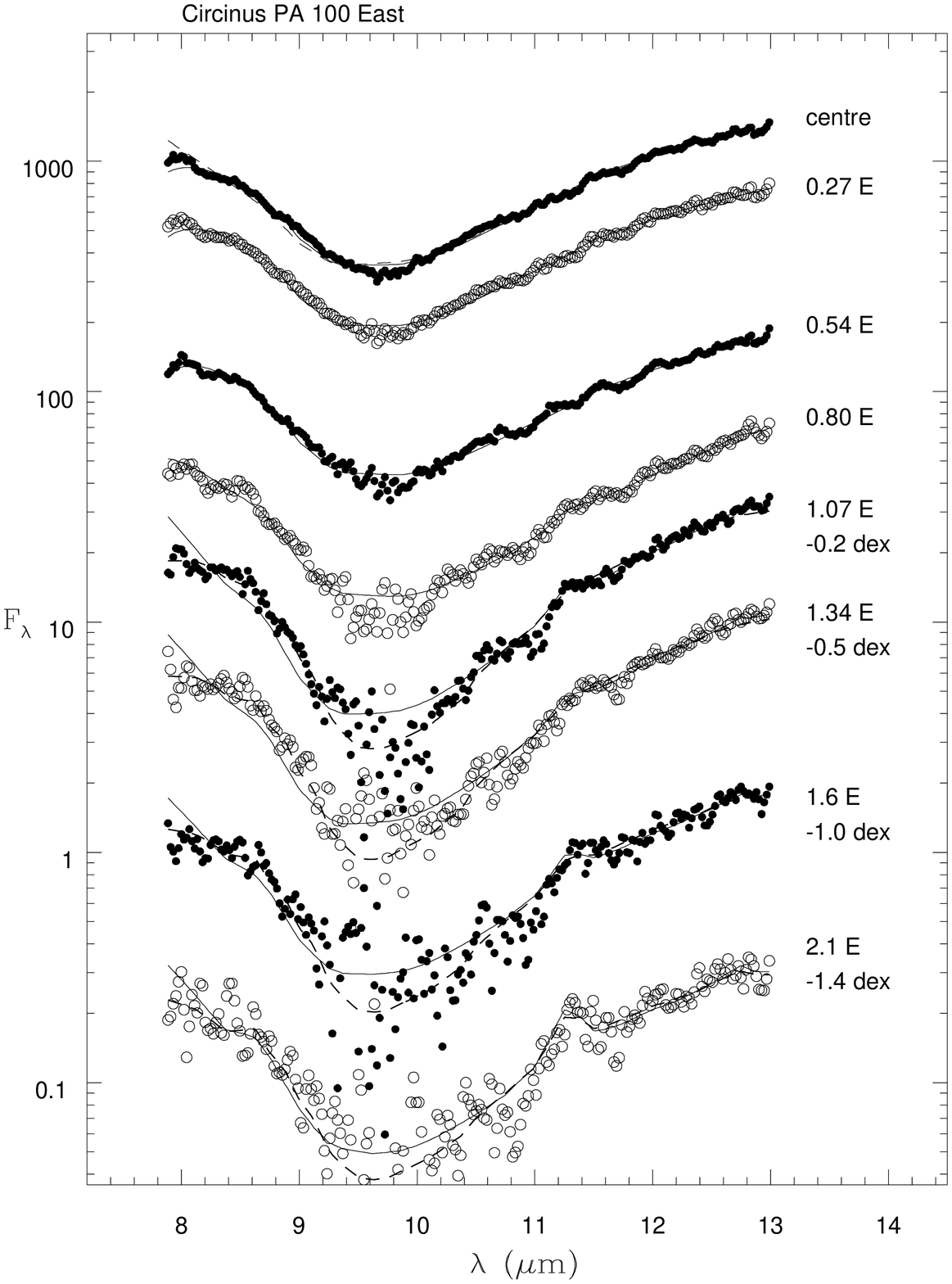}
 \includegraphics[clip=true]{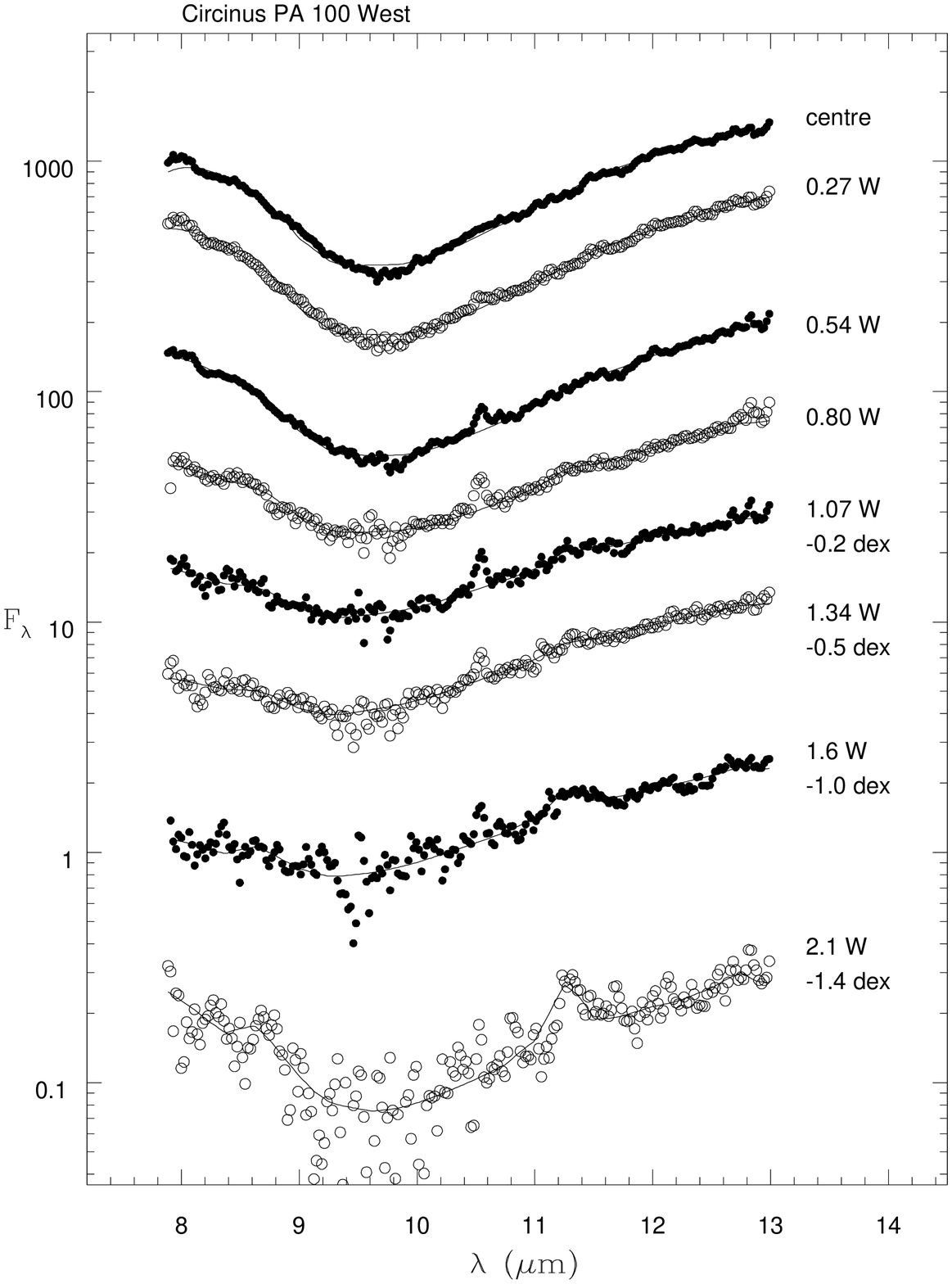}
}
     \caption{The spectra obtained on the east (left) and west (right) sides at PA 100 degrees.  The solid line shows the best fits with the Trapezium silicate emissivity function as in figure 2.  Fits with the $\mu$ Cep absorption profile are shown as dashed lines for the positions between 1 and 2 arcsec E of the nucleus  in the left panel. 
 }    
            \label{spec100}
    \end{figure*}

Long slit  spectra between 8 and 13~$\umu$m were obtained at the 8-m Gemini South telescope in clearing skies with the facility mid-infrared imager/spectrometer, T-ReCS (Telesco et al 1998), in May 2004 under programme GS-2004A-C-2.  Acquisition images at 12~$\umu$m showed structure similar to that presented by Packham et al (2005), and slit position angles of 100 and 10 degrees (North through East), along and orthogonal to the extended emission,  were selected for the spectroscopic observations (see Fig 1).  Spectra were obtained with standard chop and nod techniques (chop throw 15 arcsec) after centering the compact nucleus in the 0.36 arcsec wide slit.  The instrument was configured with the low resolution (11 line/mm) grating giving a dispersion of 0.0223 $\umu$m/pixel and a spectral resolution of 0.08$\umu$m. T-ReCS has a detector scale of 0.089 arcsec per pixel which provides a 25 arcsec long slit in the spatial direction and coverage of the full N photometric band, limited by the N filter bandpass, in the dispersion direction.  
Series of A B A B nod positions were accumulated, with total on-source integrations of 20 and 15 minutes at PA 100 and 10 degrees respectively.  The chop direction was perpendicular to the slit, and so only the signal from the guided beam position was accumulated.  A spectrum of the nearby bright star alpha Cen was obtained between the  exposures of Circinus at the different position angles.  Wavelength calibration is with respect to the planetary nebula NGC 6302, which has a rich emission line spectrum (see Casassus, Barlow \& Roche  (2000) for the spectrum and accurate wavelengths). 

The spectra were straightened, registered in the dispersion direction with reference to the sharp structure in the absorption band due to atmospheric ozone at 9.6 microns, and the dispersion calibration from NGC 6302 was applied. The T-ReCS detector displays cross talk across the different readout amplifiers as is common to this kind of detector (e.g. Sako et al 2003), resulting in smearing of the spectrum in the spectral direction.  The detector also suffered from periodic electronic pick-up noise which manifested as striping along the detector rows. To provide partial compensation for these effects, the detector signal in the unexposed region longwards of  the filter cut-off was extracted and applied to the exposed detector section. The spectra were corrected for atmospheric absorption by straightening and growing the alpha Cen spectrum along the slit, dividing it into the Circinus spectra and multiplying by a blackbody at a temperature of 5770~K (Cohen et al 1996).  The bright core of Circinus is compact and so the slit losses for the galaxy and reference star are probably similar. Nonetheless,  the calibration uncertainty is likely to be significant and we estimate $\sim$15\%, typical of mid-infrared flux calibration from the ground.  In fact the flux integrated along the slit is within 10\% of that measured in a 4.2 arcsec aperture by Roche et al (1991), confirming the compact nature of the nuclear source.

Spectra have been extracted by binning 3 rows together ( i.e in spatial increments of 0.27 arcsec, close to the diffraction limit at 8 microns) out to radii of 1.6 arcsec and 8 rows beyond.  The spectra are presented  for the PA 10 and PA 100 slit directions in figures 2 and 3 respectively.  The noise in the spectra increases between 9 and 10~$\umu$m because of  low atmospheric transmission due to absorption by ozone in the upper atmosphere, while the fluxes at these wavelengths are decreased by silicate absorption in the source.  These effects become increasingly apparent in the spatial positions furthest from the nucleus. 
Spatial cuts along the spectra have also been extracted for the two position angles by binning in the spectral dimension over $\Delta\lambda$ =0.5~$\umu$m.  These cuts have effective wavelengths of 8.2, 10.2, 11.3 and 12.5 microns respectively and are shown in fig 4, along with similar cuts for the reference star alpha Cen. For alpha Cen, low level structure seen on the right side of the cuts corresponds to the expected positions of diffraction rings; the first diffraction ring peaks at radii of 0.3, 0.4, 0.44, 0.5 arcsec for the four wavelengths shown and the 2nd and 3rd rings can also be identified.  Structure from these diffraction rings can also be discerned in the cuts of Circinus at 12.5~$\umu$m at PA 10$^\circ$, but they are more diffuse; the weak extended emission masks the diffraction structure at PA 100$^\circ$.

\section[]{Results}

We do not have accurate astrometric reference positions for the T-ReCS data, but the strong symmetry and similarity to the coronal line structure seen with the HST and VLT by Maiolino et al (1998) and Prieto et al (2004) leads us to assume that strong peak seen in the mid-infrared images is coincident with the galaxy core as argued by Packham et al. (2005).  The nuclear spectrum is dominated by a prominent silicate absorption band which produces a pronounced minimum in the spectrum at 9.7 microns, while emission bands at 11.3, 12.7, 8.6 and the wing of the band at 7.7~$\umu$m,  attributed to emission from polycyclic aromatic hydrocarbon (PAH) molecules (Leger \& Puget 1984, Allamandola, Tielens \& Barker 1985),  become increasingly prominent away from the core, consistent with previous observations (Roche et al 1991, Moorwood et al 1996). 

We have examined the spatial structure of the emission along both slit positions using the spectrum of alpha Cen as a reference (figure 4). The stellar profile is asymmetric, displaying clear structure from the subsidiary diffraction ring maxima on one side, but little evidence of them on the other, suggesting some residual aberrations in the active optics corrections.  The alpha Cen profiles are broader than the diffraction limit suggesting a significant contribution from atmospheric seeing. Allowing for the calculated diffraction-limited width from the 7.9-m clear aperture of the Gemini primary indicates that the seeing contribution was about 0.3 arcsec at 10~$\umu$m, but we can expect this to vary on relative short timescales.   We do not have any independent contemporaneous measures of the seeing at Cerro Pachon.  

To investigate the nucleus at the highest spatial resolution, we have used the last block of integrations at position angle 100 degrees and the first at 10 degrees (i.e. those bracketing the observations of alpha Cen).  The integrations amount to 5 minutes on-source, but elapsed times of about 20 minutes, and had mean intervals of 40 minutes before and 30 minutes after the observations of $\alpha$ Cen.    The Circinus profiles at both position angles have measured FWHM of 0.51 and 0.46 arcsec at 8 and 12.5~$\umu$m respectively compared to 0.39 and 0.42 arcsec for $\alpha$ Cen.  By subtracting these values in quadrature,  this suggests that the Circinus core is marginally resolved with sizes of ~0.33 and 0.2 arcsec at the two wavelengths, but we note that we would not expect a single dusty structure around the nucleus to be more extended at 8~$\umu$m than at 12~$\umu$m.  The formal uncertainties on these size estimates are only about 0.05 arcsec but seeing variations are likely to provide systematic uncertainties that exceed this. Given the unstable atmospheric conditions, it seems safer to treat these as setting an upper limit of about 0.3 arcsec on the size of the compact core of the Circinus nucleus at 8 and 12.5~$\umu$m.   From their mid-infrared images, Packham et al (2005) place a limit of $\le$0.20 arcsec diameter for the bright compact core.  

To investigate the faint extended structure, all of the accumulated data are required, and these cuts are shown in figure 4.  Here the longer integrations on the galaxy will inevitably have degraded the achieved resolution in the core.  In this figure and for subsequent spatial profiles, the negative offsets at PA 10$^\circ$ refer to positions south of the nucleus, while positive offsets at PA 100$^\circ$ refer to positions east of the nucleus. 

At PA 10$^\circ$, the flux falls rapidly away from the nucleus to the north and south, but it falls off significantly more slowly than in the stellar profile. The Circinus profiles are fairly symmetric, and on both sides, the mid-infrared spectrum shows clear PAH band emission beyond 0.8 arcsec from the nucleus. 
The PAH emission bands become increasingly prominent, accompanied by 12.8~$\mu$m [NeII] line emission, as the distance from the nucleus increases.

\begin{figure*}
   \centering
   \resizebox{\hsize}{!}{\includegraphics[clip=true]{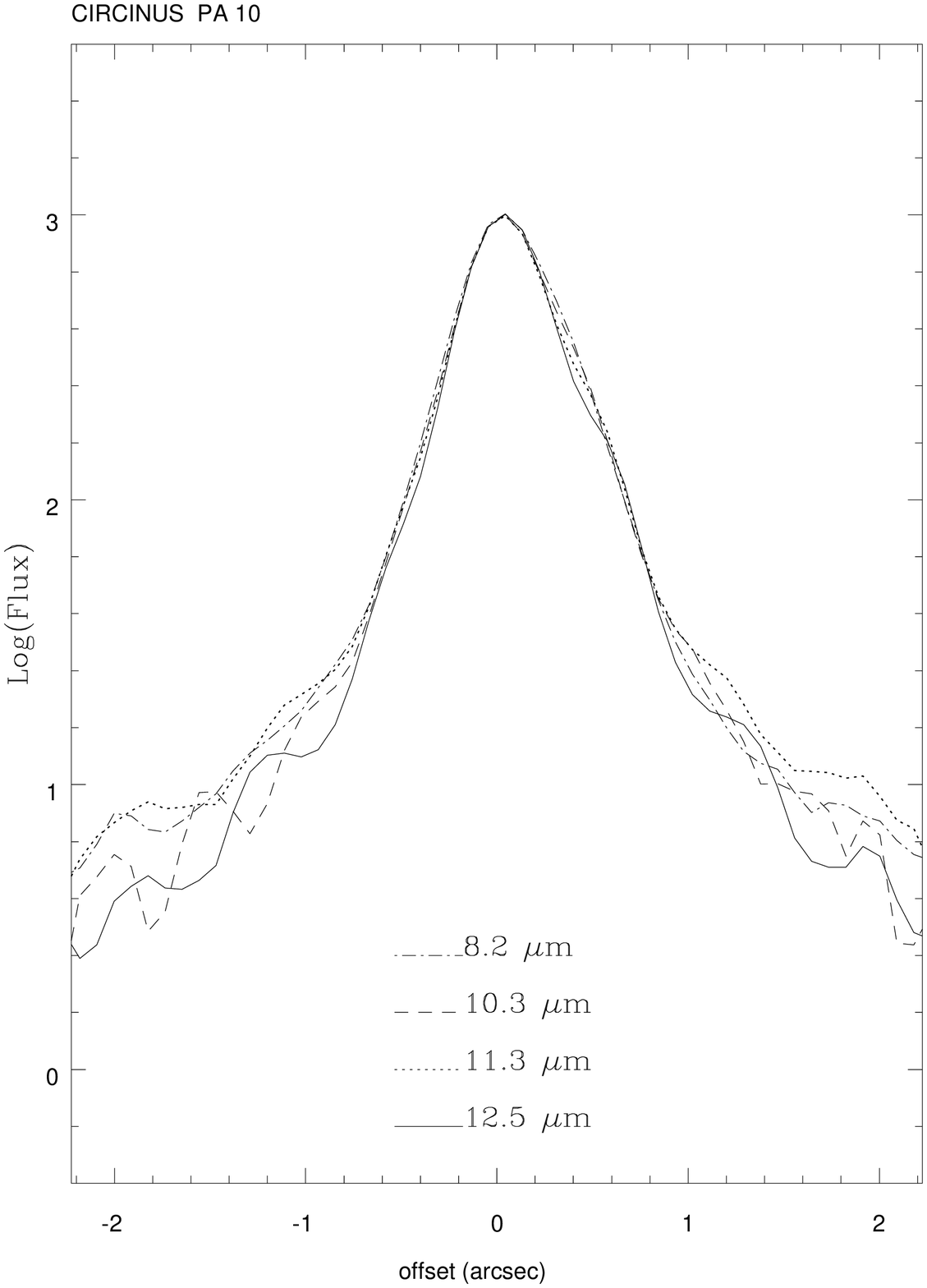}
\includegraphics[clip=true]{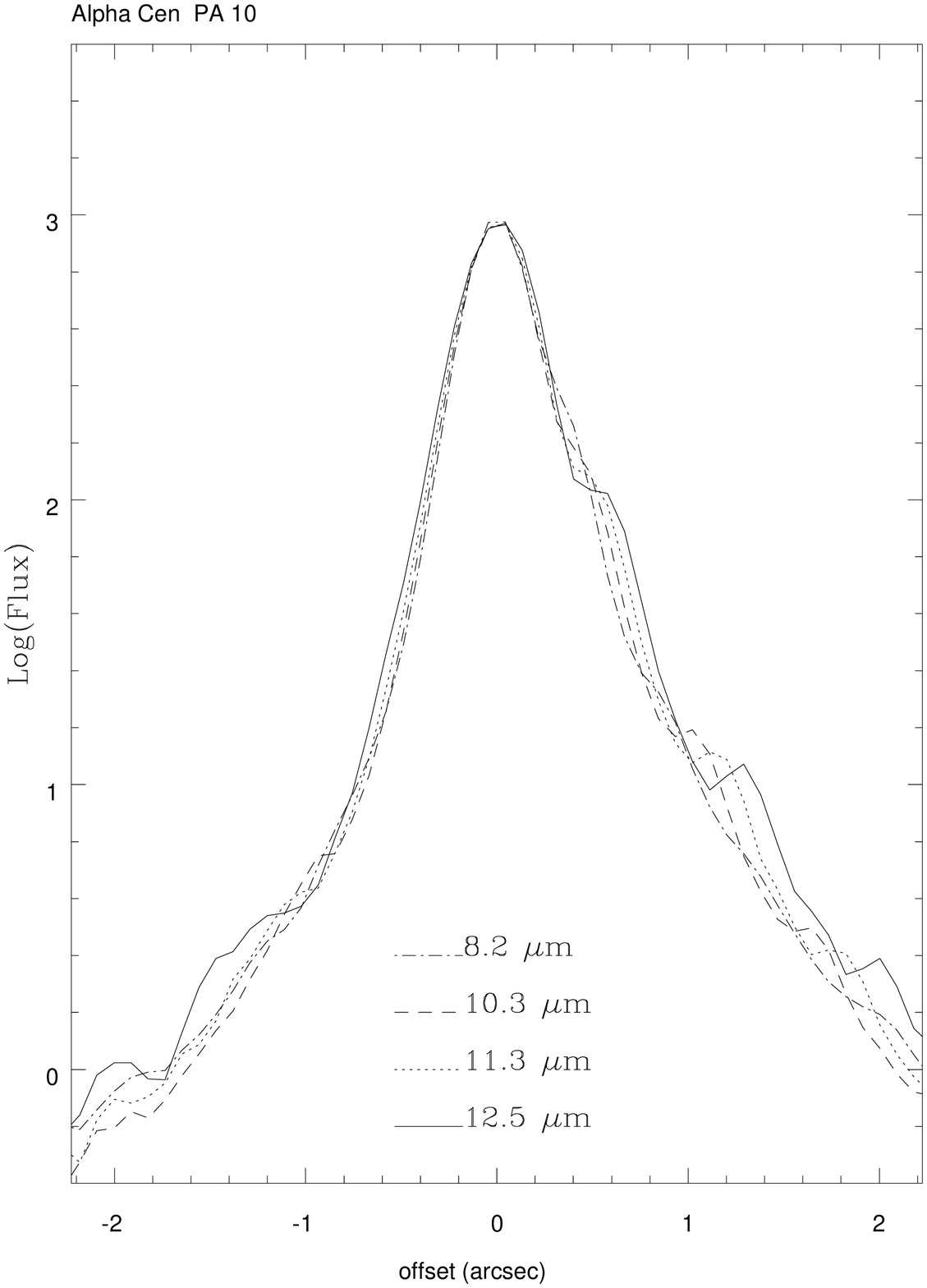}
\includegraphics[clip=true]{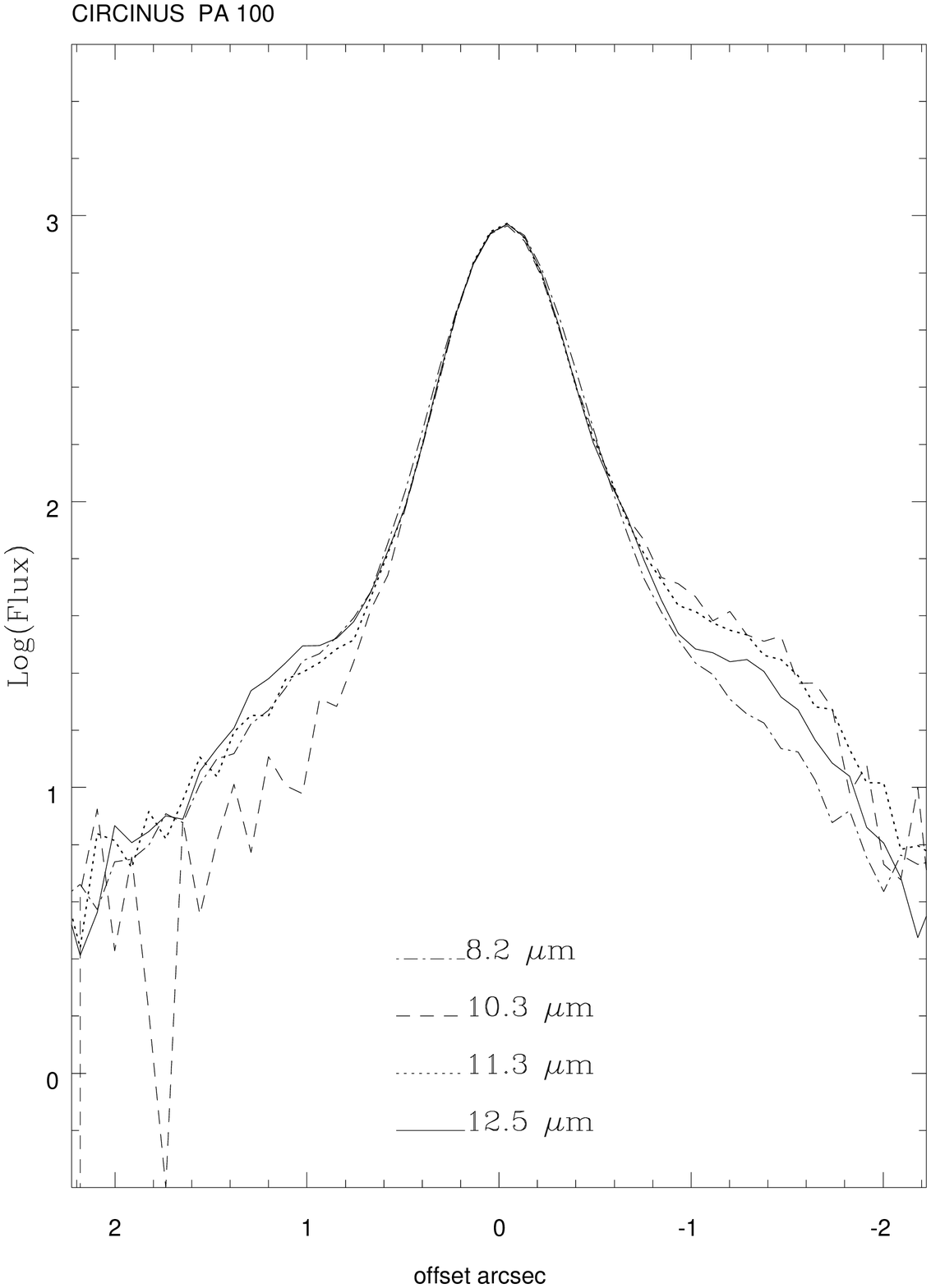}

}
     \caption{Logarithmic spatial cuts  of Circinus at PA 10 and 100 degrees compared to  alpha Cen.  Note the structure on the right edge of the alpha Cen profile attributed to diffraction rings, the approximate symmetry of the low level structure at PA 10$^\circ$ and the low level extended emission and clear asymmetry in the 10.3~$\umu$m profile at PA 100$^\circ$ in Circinus. The profiles are all normalised to the same peak flux, and use all of the accumulated data so that the spatial resoluion in the core is somewhat degraded compared to the profiles used to investigate the width of the bright core (see text). 
}  
        \label{cuts}
    \end{figure*}

The profiles at a position angle of 100 degrees show a more complex and asymmetric structure. The central core has a profile quite similar to that at PA 10$^\circ$, but the profile flattens to both the east and west beyond $\sim$0.6 arcsec.  All of the profiles show some asymmetry, but the 10.3~$\umu$m profile is markedly asymmetric, falling much more rapidly on the east side than on the west side, and suggesting a deeper minimum near 10~$\umu$m on the east side.  The spectra show that PAH emission bands become prominent about 2 arcsec from the nucleus, again approximately equidistant to the east and west, but there is little evidence of [NeII] emission.  A strong emission line of [S IV] at 10.52~$\umu$m is present on the west side of the nucleus from about 0.2 to 1.5 arcsec from the nucleus, but the line is undetected on the east side with the possible exception of the positions about 1.5 arcsec to the east.  The depth of the silicate absorption decreases with distance from the nucleus on the west side (note that the apparent increase in the depth of the minimum at 2.1 arcsec west almost certainly results from strong PAH emission structure), but remains large and perhaps increases slightly with distance from the nucleus on the east side, before decreasing again beyond about 1.5 arcsec. The colour temperature of the emission at 8 and 13 microns is about 250K at the position of the nucleus, but increases with increasing extinction to the east and decreases to the west.  \\

\begin{center}
Table 1.  Line and emission band fluxes in the extended emission regions. 

\hspace*{3 mm}

\begin{tabular} {l l l l}
		 & [SIV] & [NeII] & 11.3$\umu$m \\[2 mm]
PA 10$^\circ$ & 2.0$\pm$1 & 10$\pm$ 3 & 21$\pm$4 \\
PA 100$^\circ$ & 4.7$\pm$1 & 4.1$\pm$ 3 & 15$\pm$5  \\
ISO                  & 13 & 96 & 390  \\

\end{tabular}
\end{center}
\noindent 
Note: \\
Intensities are summed along the inner 5 arcsec of the slits, but omitting the central 0.6 arcsec, and are in units of 10$^{-20}~$Wcm$^{-2}$ \\[5mm]

While emission lines of [SIV] and [NeII]  and PAH dust band emission are clearly present in many of the off-nucleus spectra, these features have much lower equivalent width at the central positions within 0.5 arscec of the nucleus. Because the compact core accounts for more than 90\% of the total flux, uncertainties in the nuclear spectrum will dominate estimates of the total intensities of these features.  We have therefore integrated along the central 5 arcsec of the slits to estimate the intensities of these feature but have excluded the data from the central 0.6 arcsecond, so that these values correspond to the extended arcsecond scale emission outside the nucleus.  The line and 11.3~$\umu$m PAH band  intensities at the two position angles are  compared to the large aperture ISO measurements of Moorwood et al (1996) in Table 1.  The fluxes are  estimated by integrating over the expected line positions and subtracting a local linear continuum.  More than one third of the [SIV] line emission measured by ISO falls within the T-ReCS slit at PA 100$^\circ$ while less than 10\% of the [NeII] flux is detected, suggesting that the [SIV] emission is largely confined to the coronal line region, while the [NeII] emission is more extended.  The [NeII] intensity at PA 10$^\circ$ is twice that detected at PA 100$^\circ$.  The 11.3~$\umu$m band is much more extended, with only $\sim$5\% falling within the slit; the bulk of the PAH band emission presumably  arising in the circumnuclear star forming regions (e.g. Elmouttie et al 1998). 

\subsection{Fits to the spectra}

In order to quantify the behaviour of the absorption, we have fit the spectra with different emission and absorption components.  The fits cannot be physically realistic in this complex region, but do allow us to extract some quantitative data. 

Following Aitken \& Roche (1982) we have fit the spectra with emission spectra suffering extinction by cool silicate grains.
The emission spectra consist of black-body and silicate grains together with a PAH emission spectrum extracted from the Orion ionisation front (Roche 1989).  We employed silicate grain profiles derived from the Trapezium in Orion and the M supergiant $\mu$ Cephei, which are taken to be representative of molecular clouds  and the galactic interstellar medium and oxygen-rich circumstellar environments respectively (e.g. Roche \& Aitken 1984).  The redshift produced by the recession velocity of 439 km~s$^{-1}$ (Freeman et al 1977) is taken into account in the fits. 

  The goodness of fit is estimated from reduced $\chi^2$ values, with the errors estimated from the scatter in the data points. These allow us to differentiate between different combinations, but the range of values giving  adequate fits can be quite large for some  spectra.   In particular, the depth of the silicate absorption depends critically on the assumed spectral properties of the underlying emission. 
  
Whilst these fits are not unique, they do indicate that the fits improve with a contribution from warm emitting silicate grains in the central nucleus.   For example, the best fits to the central spectra with featureless and Trapezium silicate grain emission together with silicate absorption have  $\chi^2$/N= 1.9 whereas omitting the silicate grain emission results in a value of  $\chi^2$/N= 3.0.  Inspection of these fits shows that the main difference between the fits with and without emitting silicate grains occurs in matching the short wavelength edge of the silicate absorption, but that the overall fit is not qualitatively different despite a significant difference in the absorption depth.    The best fits are shown in figures 2 and 3.   Any contribution from  the warm emitting silicate grains decreases with distance from the nucleus and a silicate emission component does not improve the fit beyond 1 arcsec.    In most spectra, the silicate profile is better fit by grains with an emissivity similar to the Trapezium rather than $\mu$ Cep.  The only exception to this is at about 1.5 arcsec east of the nucleus where the minimum at 9.7~$\mu$m is sharper than  the Trapezium profile and the $\mu$ Cep emissivity function gives a better match; at the position 1.6 arcsec E, the values of $\chi^2$/N for fits with featureless grains plus the PAH emission spectrum overlaid by cool silicates are 3.3 and 2.5 with Trapezium and $\mu$ Cep emissivity functions respectively.   This could indicate that the cool absorbing dust producing the asymmetry between the east and west spectra has a sharper emissivity function than the Trapezium, but definite conclusions are not warranted given the simplicity of the models used here. 

Nonetheless, with these simple models of a warm emitting source and a cool absorbing screen, the fits give information on the silicate profile and indicative estimates of the apparent absorption optical depth.  They also allow us to make estimates of the contribution of the PAH emission bands to the total emission.  

\begin{figure*}
   \centering
   \resizebox{\hsize}{!}{\includegraphics[clip=true]{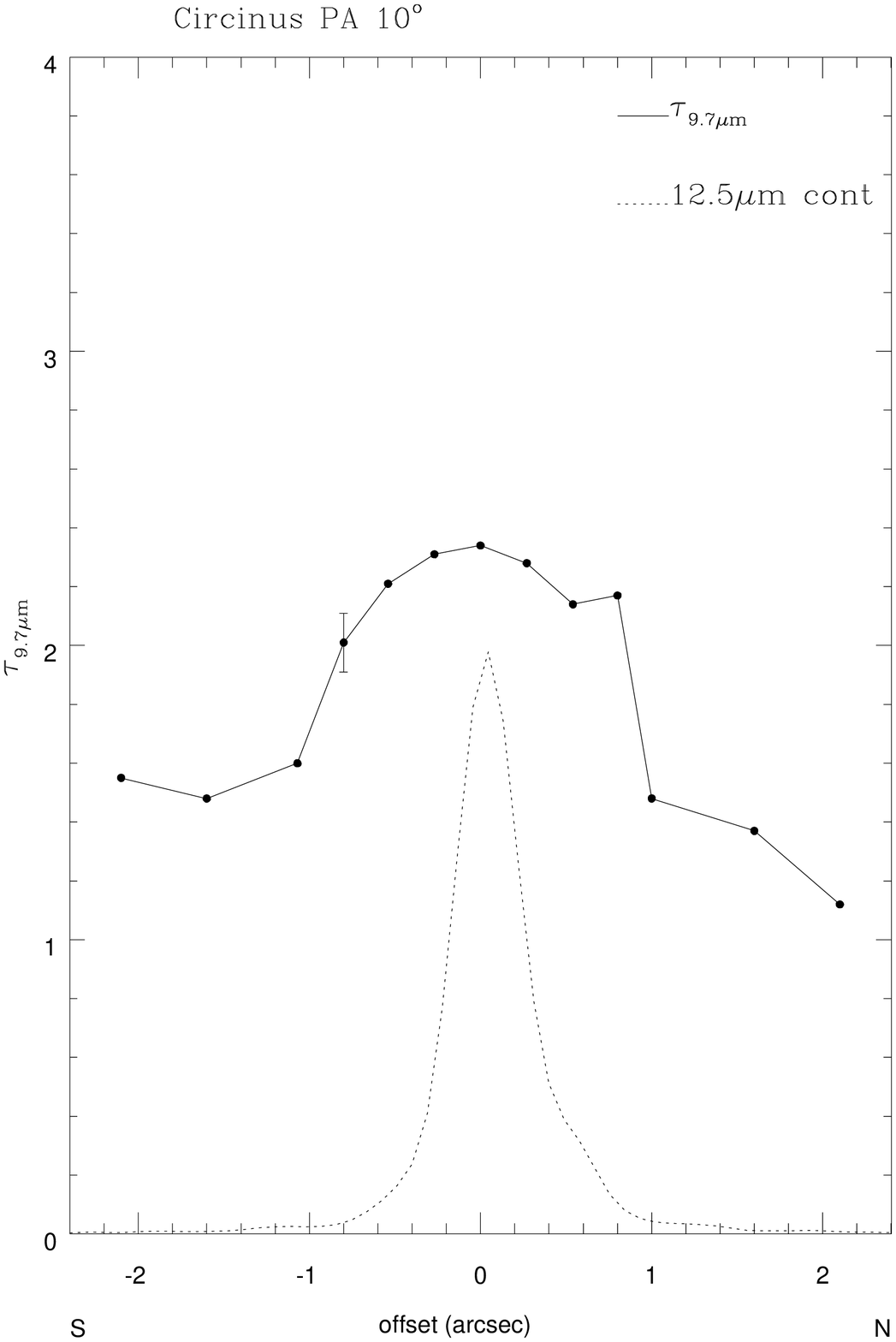}
 \includegraphics[clip=true] {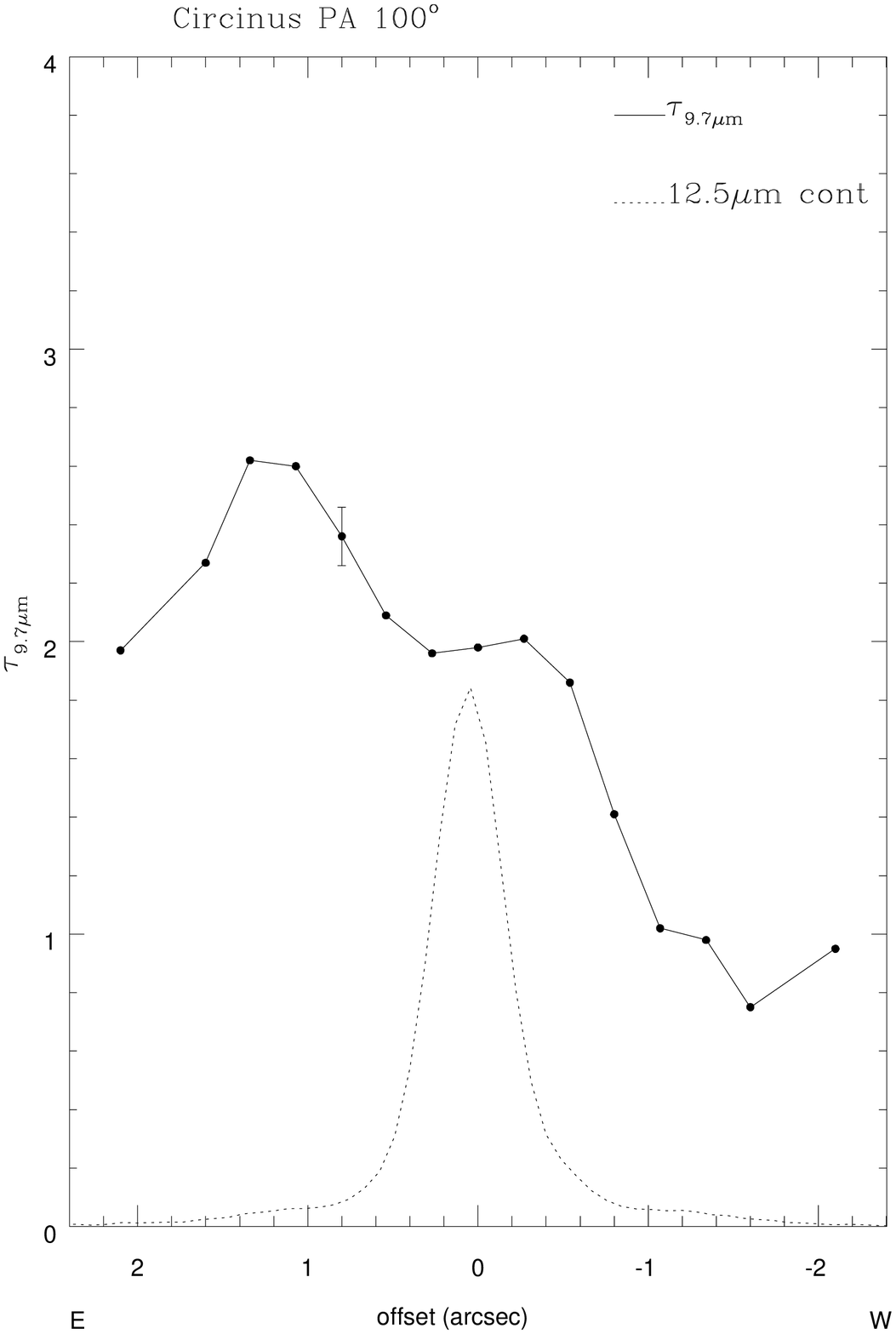}

}
     \caption{The spatial variations in the optical depth at the minimum of the silicate absorption band at 9.7~$\umu$m at PA 10$^\circ$ and 100$^\circ$,  assuming that the underlying emission may be represented by a black body spectral distribution together with PAH emission bands. Representative statistical error bars are shown, but systematic effects, including possible underlying silicate emission and the likelihood that the PAH emission bands come from a different spatial region almost certainly dominate the uncertainties.   The normalised 12.5~$\umu$m continuum profiles are shown as dotted lines.   }
        \label{tau}
    \end{figure*}

\begin{figure*}
   \centering
   \resizebox{\hsize}{!}{\includegraphics[clip=true]{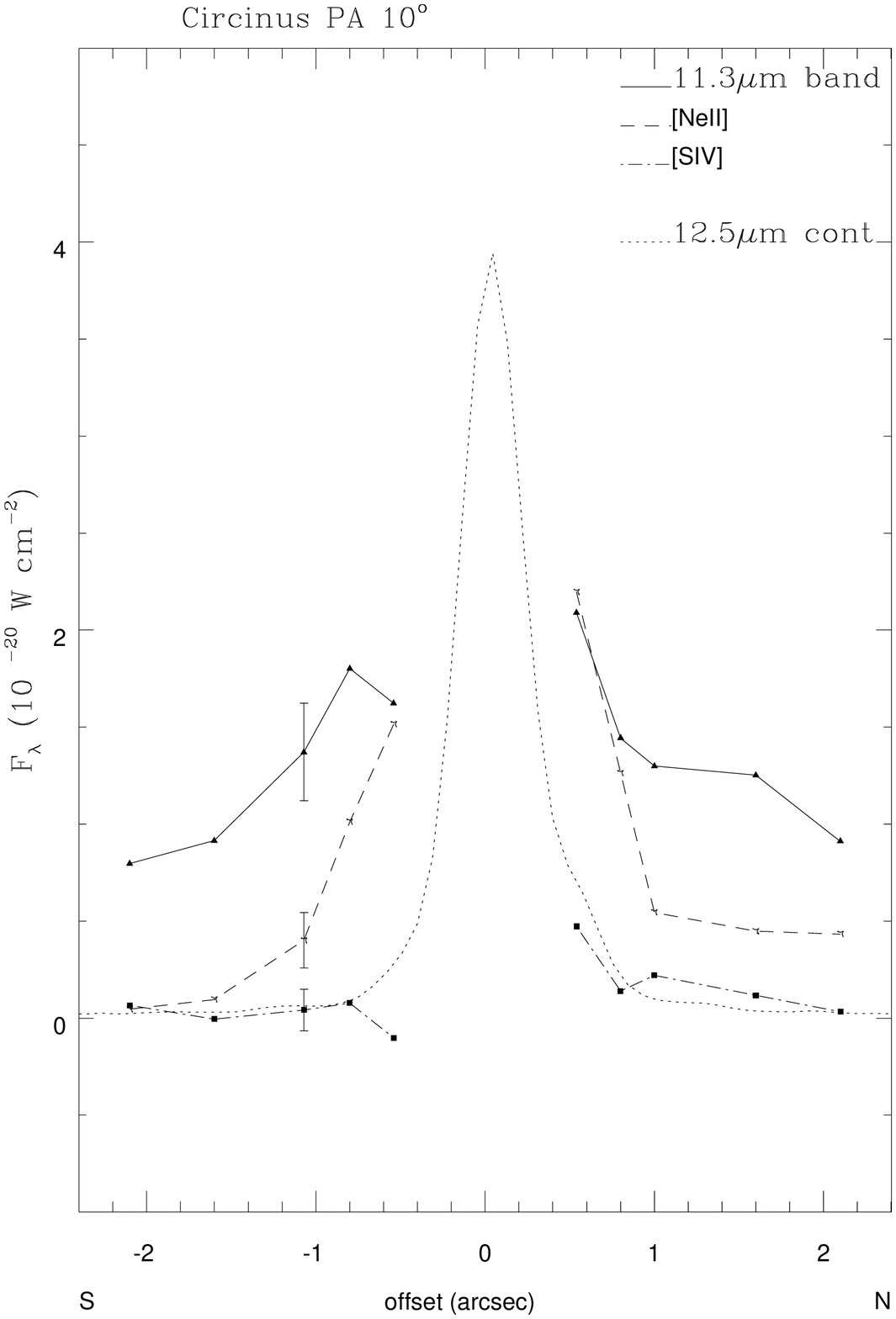}
 \includegraphics[clip=true] {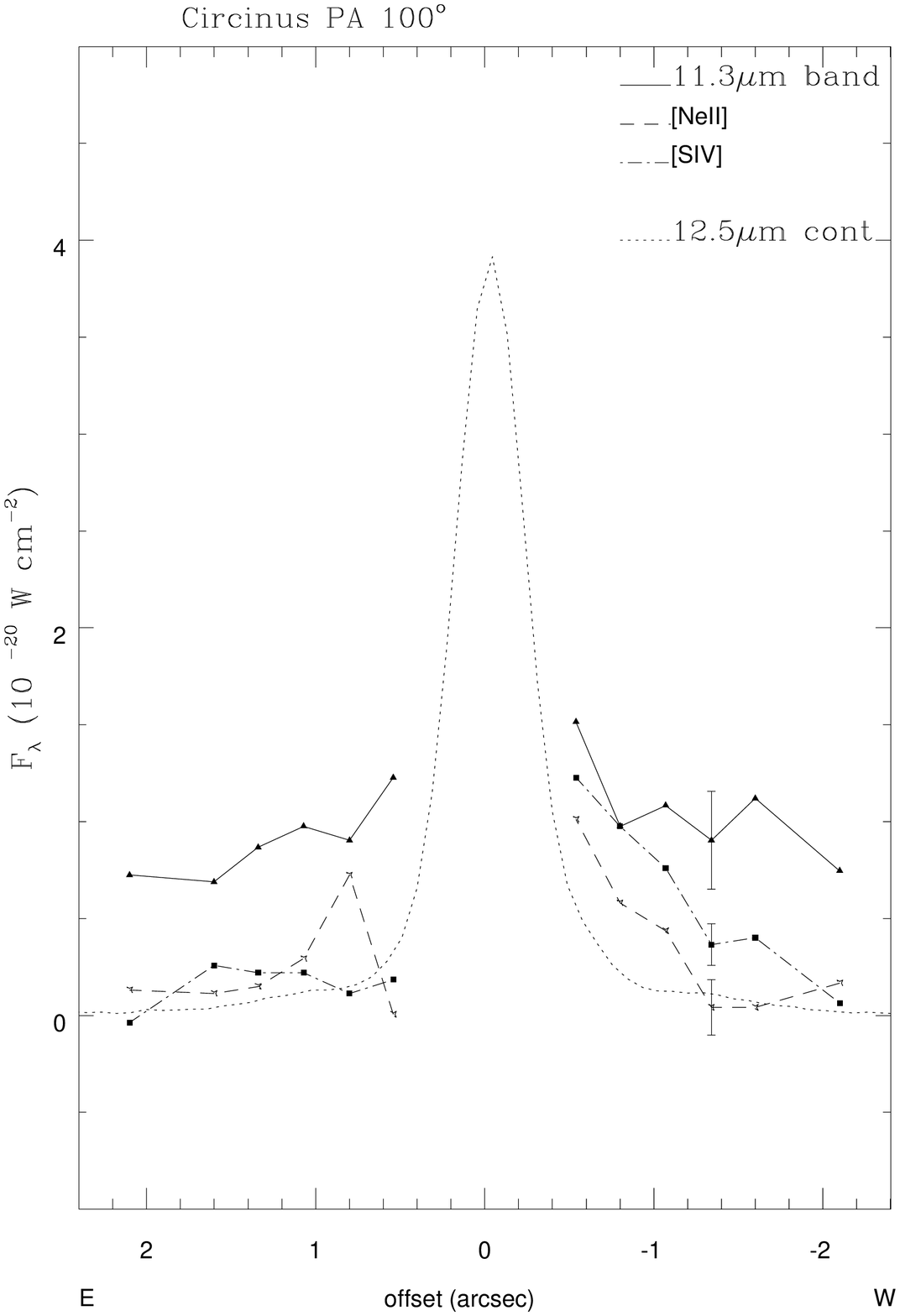}

}
     \caption{Spatial distributions of the [SIV] and [NeII] emission line and the 11.3~$\umu$m band intensities  at PA 10$^\circ$ and 100$^\circ$.  The line intensities plotted are in 0.267 arcsec bins in the spatial direction, and are extracted by subtracting linear continua extrapolated from the flux on both sides of the line positions.  Indicative error bars are shown, but note that the uncertainties increase rapidly towards the nucleus because of the bright continuum emission. The normalised 12.5~$\umu$m continuum profiles are shown as dotted lines.   }
  
        \label{lineprofiles}
    \end{figure*}

\begin{figure*}
   \centering
   \resizebox{\hsize}{!}{\includegraphics[clip=true]{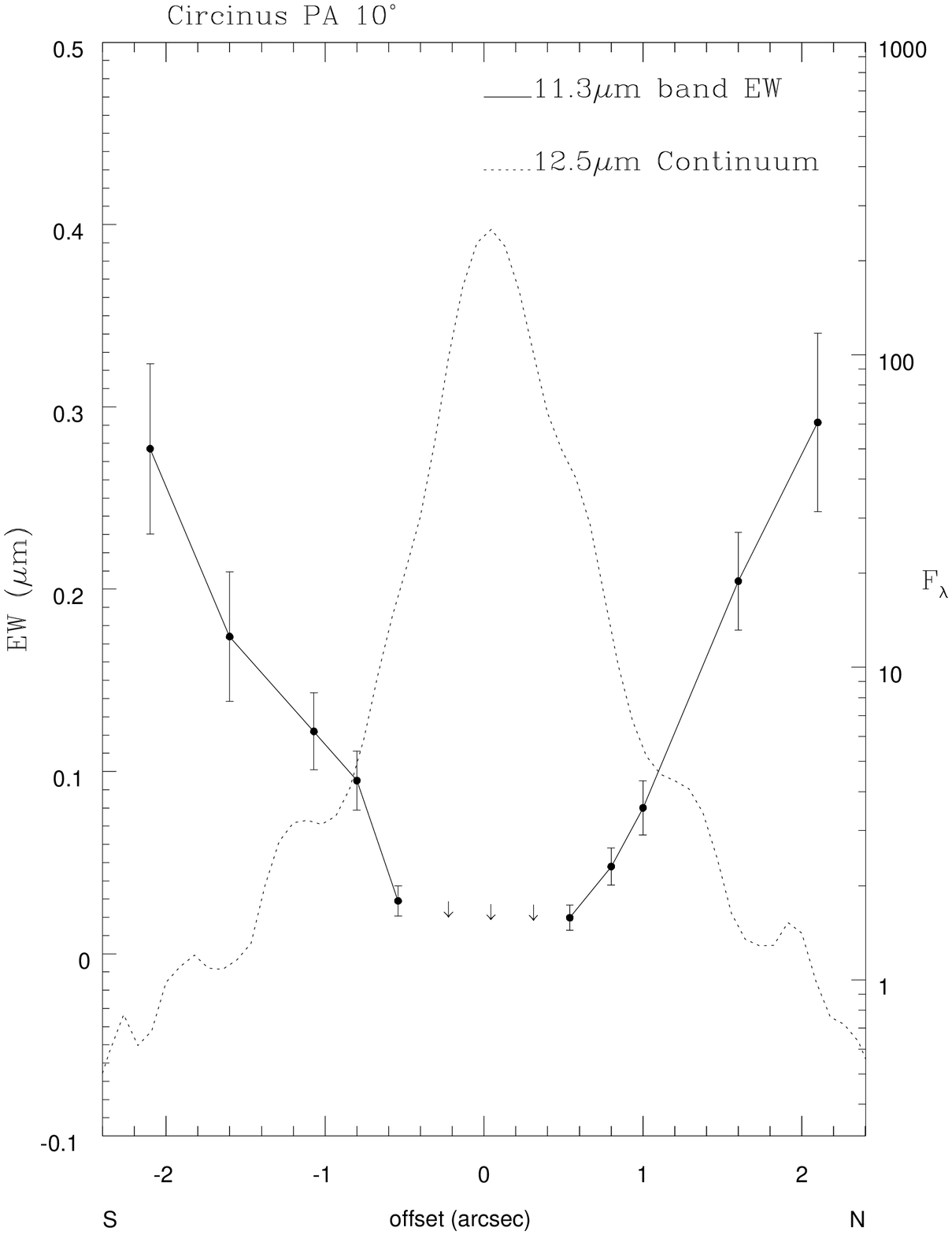}
 \includegraphics[clip=true] {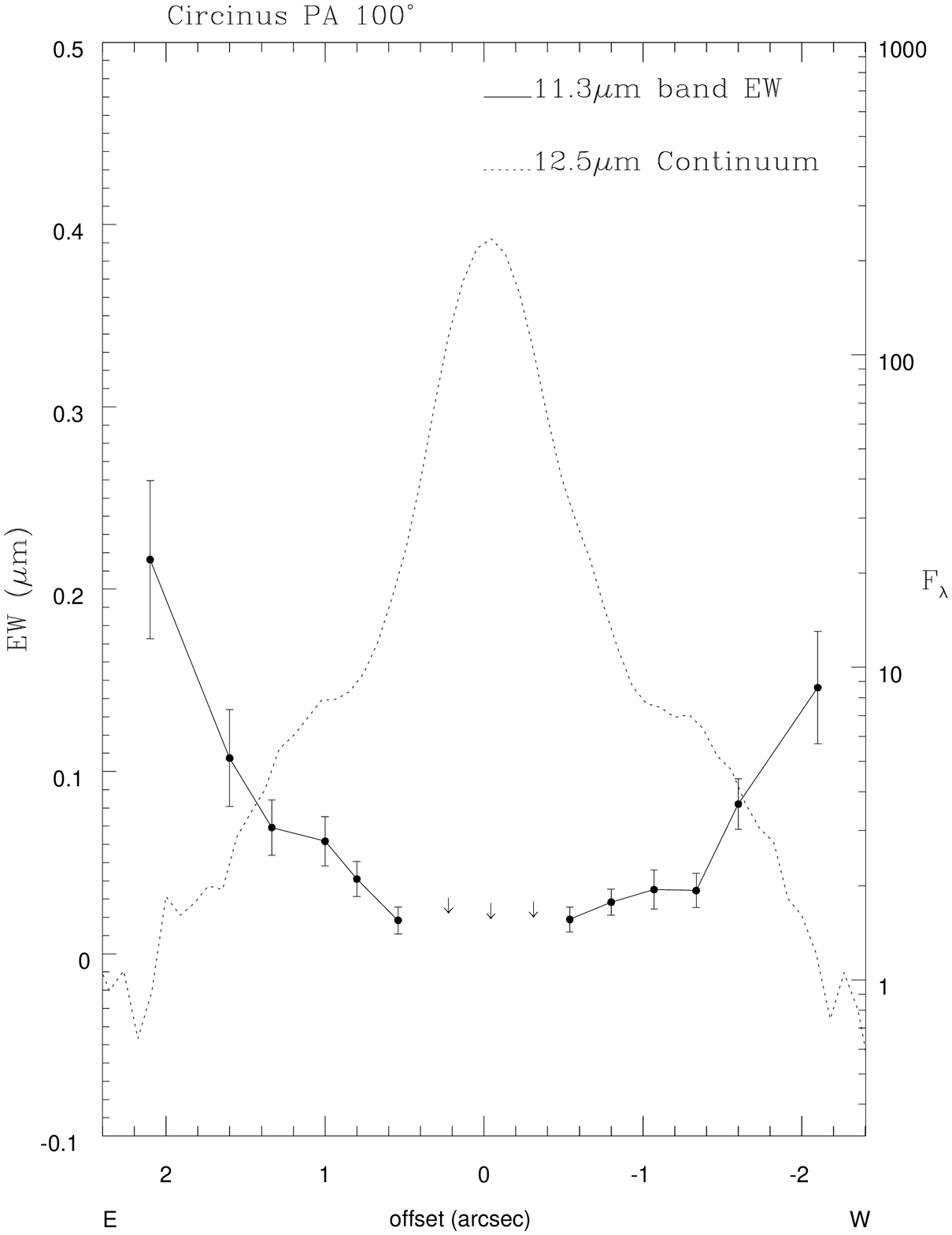}

}
     \caption{The spatial distribution of the equivalent width of the 11.3~$\umu$m dust band (solid line) compared to the log of the 12.5~$\umu$m continuum flux at PA 10$^\circ$ and 100$^\circ$.   The 11.3~$\umu$m equivalent width is in microns and refers to the left axis. The band is not detected  in the central core positions and 3~$\sigma$ upper limits to the equivalent width are shown as arrows. The 12.5~$\umu$m continuum flux is scaled so that the full dynamic range can be seen, and the scale is drawn on the right axes.}  
        \label{EW11}
    \end{figure*}

\section{Discussion}

The luminosity of the nucleus is estimated at $\sim2 \times 10^{10}$ L$_\odot$  (e.g. Oliva et al 1999). Directly heated dust grains in equilibrium with the radiation field will have temperatures $>$1000~K at distances $<$0.1pc and $\sim$250~K at distances of $\sim$1~pc. We might expect the distance to the inner edge of a circumnuclear disk or torus to be set by the evaporation of dust grains, which would correspond to the former scale, while the bulk of the emission at 10~$\umu$m might correspond to the latter.  These simple estimates are consistent with the scales estimated for the putative torus around the Circinus nucleus derived from models (e.g. Ruiz et al 2001).  Both scales lie well within the T-ReCS resolution limit, and so we therefore do not expect to resolve emission from directly-heated dust grains if they are similar to those in the Galactic interstellar medium, although the smaller grains will reach higher  temperatures. However, grains can achieve higher temperatures through several mechanisms including: disk heating, non-equilibrium emission and through heating by trapped emission line photons.

\subsection{Silicate Absorption}

Prominent silicate emission has now been detected in several active nuclei.  Roche et al (1991) identified weak silicate emission features in only 2 of their sample of 60 infrared-bright galaxy nuclei, but silicate emission has  recently been confirmed in one of those objects and detected in several more quasars and AGN by Spitzer (e.g. Weedman et al 2005).   While we might expect silicate emission to become more prominent in the hotter regions closer to the nucleus, this is also the region where small grains are likely to be preferentially destroyed, and it is not yet clear how the silicate emission detected in these luminous objects  relates to dust in a torus close to the nucleus. 
 
The fits indicate that the optical depth of the silicate absorption towards the Circinus nucleus is in the range 2.2$\le$  $\tau_{9.7\umu{\rm m}} \le$3.5 with the larger value corresponding to the case where the underlying emission spectrum contains a prominent silicate emission component, and the smaller value to underlying blackbody emission.  However, to interpret the spectra, we make the assumption that any silicate emission component is small and that the underlying spectrum can be represented by a black body spectral distribution.  This is in line with the relatively weak emission bands seen in AGN, although it is of course possible that silicate emission does arise from warm dust in the inner torus. 

For the spectra at PA 10$^\circ$, the absorption optical depth shows a roughly symmetric behaviour about the nucleus, remaining approximately constant at $\tau_{9.7\umu{\rm m}} \sim$2.2  out to $\sim$0.6 arcsec north and south of the nucleus before falling to $\tau_{9.7\umu{\rm m}} \sim $1.4 at a distance of 1.5 arcsec.  At PA 100$^\circ$, the  absorption shows marked asymmetry.  The optical depth rises with increasing distance from the nucleus to the east, reaching a maximum value  $\tau_{9.7\umu{\rm m}} \sim$2.6 at 1.4 arcsec east, before falling again.  On the west side of the nucleus, the optical depth decreases steadily, falling to $\tau_{9.7\umu{\rm m}} \sim$1.5 at 1 arcsec west and $\sim$1.0 at 1.4 arcsec west. The spatial variations of the absorption optical depth are shown in figure 5.

As a check on the values of extinction at the positions observed 2.1 arcsec from the nucleus,  where the PAH bands are most prominent, we have used the prescription described by Aitken \& Roche (1984).  This uses the observed relationships between the equivalent width of the 11.3~$\mu$m band and the ratio of the flux at 10~$\umu$m to the flux at 8.1 and 13~$\umu$m in relatively unobscured Galactic objects to estimate the depth of the silicate absorption in galaxy nuclei.  
For Circinus, we obtain estimated extinctions of $\tau_{10\umu m} =$ 1.4 and 1.3 for the positions 2.1 arcsec north and south of the nucleus and  $\tau_{10\umu m} =$ 1.6 and 1.1 for the positions 2.1 arcsec east and west of the nucleus respectively, with uncertainties of 0.2.  These are adequately close to the values obtained through the fits, but do of course rely on the same assumptions of warm emitting material suffering extinction by a cool overlaying  medium. 

We note that if an underlying silicate emission band is included in the fits, the spatial profile at  PA 10$^\circ$ displays a similar symmetry, but with a significantly greater peak optical depth, while the profile at PA 100$^\circ$ remains asymmetric. However, the inclusion of a warm  emitting silicate component at PA 100$^\circ$ increases  the optical depth towards the nucleus, but has little effect at the positions 1.5 arcsec to the east and west (these positions have lower absorption optical depths than towards the nucleus, reflecting the smaller contributions from silicate emission required in the fits). In this case, the nucleus appears to suffer a greater extinction than the extended emission, but the difference in optical depth between the positions 1.5 east and west remains similar to the case with an underlying blackbody emission component at $\Delta\tau_{9.7\umu{\rm m}} \sim$1.6.

The statistical uncertainties delivered by the fitting routines are quite small for most spectra (e.g. typically $\le0.1$ for the error in the optical depth), but these are dwarfed by potential systematic effects.
We emphasise that the conclusions from the fits are not unique; There could be multiple sources or variations in  the emitting material on the east and west sides which would affect the underlying emission spectrum and so the depth of the absorption.  However,  it seems more natural to interpret the results in terms of variations in the column of cool material along different lines of sight.  We conclude that the behaviour of the silicate absorption confirms the observations made at other wavelengths- that the east side of the nucleus suffers greater extinction than the west side (e.g. Maiolino et al 2000) - and although extraction of detailed quantitative data depends on the emissive properties of the underlying material which are not known, it does allows us to estimate relative values that allow us to probe the dust temperature structure and distribution of the circumnuclear material.

\subsection{Line Emission}

The intensities of the [SIV] and [NeII] emission lines and the 11.3~$\umu$m PAH band have been measured for each spatial position as described above and plotted together with the 12.5~$\umu$m continuum profiles in figure 6.     Note that in the core, where the continuum emission is very high, the extracted line intensities are very uncertain, and so these quantities are not plotted here, while in the outer regions the flux is low and the fractional uncertainties are correspondingly large.  Indicative error bars are shown for the points plotted.  These plots demonstrate the asymmetry in [SIV] emission on the east and west sides of the nucleus, contrasting with the fairly symmetric distribution of the 11.3~$\umu$m PAH band emission. 

The [SIV] line emission is very asymmetric, being prominent at about 1 arcsec west of the nucleus, but undetected at the corresponding position to the east.   On the west side, the line flux decreases with distance from the nucleus  despite decreasing extinction, and is undetected at 2 arcsec where the equivalent width of the 11.3~$\umu$m band increases sharply and the spectrum is dominated by PAH emission.  The line is not detected to the north or south.  This behaviour mirrors the [Si VI] and [Si VII] emission structure reported by Maiolino et al (2000) and Prieto et al (2004), and suggests that the [S IV] line is largely confined to the high excitation coronal zone. The fact that almost half of the [SIV] emission measured by ISO is contained within the 0.36 arcsec T-ReCS slit, which is narrower than the coronal emission line region mapped in the [SiVI] and [SiVII] lines, suggests that  the great majority, and perhaps all of the ISO [SIV] flux is emitted from the coronal region.  Maiolino et al (2000) and Prieto et al (2004) find that the peaks of the [Si VI] and [SiVII] line emission coincide with the nucleus. This does not appear to be the case for [SIV].  Although, the strong continuum makes extraction of the [SIV] flux uncertain right in the core, the extracted flux is lower than  at  0.5 arcsec east, and the limits indicate that the [SIV] emission is not strongly peaked.  Comparison of the available coronal line maps indicates that [SiVII] and [Al IX] (Prieto et al 2004, Maiolino et al 1998), which require $>$200~eV for photoionisation, are the most compact species, and [SIV] ($>$35~eV) the most extended, suggesting that we may be seeing ionisation stratification in the coronal zone on subarcsecond scales, as suggested by Maiolino et al (1998).  High spatial and spectral resolution spectroscopy would allow this to be investigated in more detail.    

If the underlying emitting structure is symmetric, the weak  [S IV] emission to the east suggests that this region suffers an additional absorption of a factor $\ge$4 compared to the west side, consistent with the additional silicate optical depth.

Oliva et al (1999) present results from a grid of ionization models to investigate element abundances and density structures in Circinus.  With the available observations, they were not able to determine the spectral shape of the ionising continuum, but did conclude that dust grains are mixed  with the gas in the low excitation region.  We see direct evidence for the survival of dust grains in the high excitation region.  The spectra at 1--1.5 arcsec east show mid-infrared continuum emission at a colour temperature of about 200K after allowing for the silicate absorption, with a relatively small contribution from the PAH emission bands and little evidence of silicate emission; this continuum must arise from relatively warm dust grains, presumably mixed in with the gas in the coronal zone.  Survival of dust grains in these regions is perhaps surprising, but there is evidence that grains can survive in such hostile environments.  For example, Casassus et al (2000) find that some refractory species are heavily depleted in the coronal zone of the planetary nebula NGC 6302,  suggesting that grains coexist with very high excitation gas. These grains in the coronal zone are likely to have different compositions from dust in the Galactic interstellar medium, but will absorb some of the trapped UV photons, reducing the extent of the coronal zone.  The relatively featureless underlying dust emission spectra, with very weak or absent silicate or PAH emission structure suggests that the emitting grains are not the same as the small  grains in the interstellar medium, but could be large or of different composition (e.g. see Laor \& Draine 1993).

\subsection{PAH Emission bands}

It is well established from  studies of regions such as the Orion nebula (Aitken et al 1979, Sellgren 1981) and the planetary nebula NGC 7027 (Aitken \& Roche 1983), that the mid-infrared PAH emission bands peak just outside the ionised regions in Galactic nebulae, suggesting that the carriers of these bands do not survive within HII regions.  The PAH bands are believed to arise through emission from small hydrocarbon grains excited by single photon absorption (Sellgren 1984).   Aitken \& Roche (1985) argued that the small hydrocarbon grains are destroyed by the hard photon flux from AGN, accounting for the marked differences between the mid-infrared spectra of galaxies with active nuclei and those dominated by star-formation.  This has since been quantified (e.g. Voit 1992) and the observational evidence extended (e.g. Genzel 1998).  In Circinus we have the opportunity to investigate the nuclear region in detail.   

As shown in figure 6, the 11.3~$\umu$m band emission has a markedly different distribution from the mid-infrared continuum emission and the line emission. This is shown more clearly in figure 7 where the spatial distributions of the equivalent width of the 11.3~$\umu$m band are plotted along with the logarithm of the 12.5~$\umu$m continuum flux;  the extinction at 12.5~$\umu$m is relatively low and plotting the logarithmic flux emphasises the low level emission structure.  At PA 10$^\circ$, the 11.3~$\umu$m equivalent width is low near the nucleus, but increases rapidly outside the central 2 arcsec. At PA 100$^\circ$, the equivalent width remains low out to $\sim$1.5 arcsec from the nucleus and then increases, with the region of low equivalent width coincident with the low-level extended emission to the east and west of the nucleus where the coronal line emission is detected.  The equivalent widths at distances of  2 arcsec from the nucleus have reached values between 0.16 and 0.3~$\umu$m, and may increase further at larger distances.   These values are at the lower end of those seen in typical HII region galaxies (e.g. Aitken \& Roche 1984). 

The slower increase in 11.3~$\umu$m band equivalent width with separation from the nucleus at position angle 100$^\circ$ than at 10$^\circ$ suggests that the PAH molecules are depleted in the regions close to the coronal line emitting regions.  It is especially noteworthy that in the spectra obtained at PA 100$^\circ$, the 11.3~$\umu$m band intensity is approximately equal on the east and west sides at distances between 1 and 1.5 arcsec from the nucleus, despite a very large difference in the estimated extinction ($\tau_{9.7\umu{\rm m}} \sim$2.5 and $\sim$1.0 respectively) and correspondingly large differences in the [S IV] line flux.  This is apparent in the 11.3~$\umu$m band spatial profile in figure 6 while the decreased continuum emission resulting from the increased absorption accounts for the asymmetry in the spatial profile of the 11.3~$\umu$m equivalent width at PA 100$^\circ$ in figure 7. This suggests that the 11.3~$\umu$m band emission arises predominantly outside the regions affected by the large columns producing the absorption and further from the nucleus, while the line emission arises closer in and behind the obscuring material on the east side.

\subsection{Interpretation}

 The absorption column inferred from X-ray measurements is more than a factor of 10 greater than that implied by the silicate band, consistent with the X-ray emitting zone being much closer to the active nucleus than the dust emitting at 10~$\umu$m.  The strong variations in the depth of the silicate absorption on subarcsecond scales indicate that the cold absorbing dust must be fairly local to the nucleus on scales of tens of parsecs.  From observations of water masers, Greenhill et al (2003) have inferred the presence of a disk of molecular material inclined by about 65$^\circ$ to the line of sight and tilted so that the eastern side of the disk lies towards the earth.  If  the mid-infrared absorption arises in an inclined structure with a similar orientation, it would naturally explain the east-west asymmetry, although the scale measured by the water masers is $<$1pc compared to tens of parsecs at 10~$\umu$m.  The additional optical depth between the east and west sides, $\tau_{9.7\umu{\rm m}} \sim$1.6, corresponds to A$_V \sim$25 mag (Roche \& Aitken 1984) or a column density of n$_H \sim 5 \times 10^{22}$ cm$^{-2}$ if the absorption arises in grains similar to those in the Galactic interstellar medium.  Integrating this over a disk radius of 30 pc, and allowing for the inclination gives a total mass of $\sim 4 \times 10^5$ M$_\odot$ for a uniform column density.  This compares to an upper limit of  $4 \times 10^5$ M$_\odot$ for the mass of the disk inside 0.4pc  derived by Greenhill et al (2003) for the molecular disk containing the water masers.  The structure detected in the mid-infrared is much less dense than the inner disk, but still has a very substantial optical thickness at a radius of 30~pc, and could have a much greater value closer to the nucleus.
 
It is intriguing to note that the east side of the nucleus appears to suffer increased extinction on a range of spatial scales.  Inspection of the H$\alpha$ images taken with the Hubble Space telescope and presented by Wilson et al (2000) suggests that a significant part of the structure visible may be due to extinction variations.  In particular, the deep image shown in figure 12 of Wilson et al (2000) is suggestive of a dust lane oriented approximately North-South, which in turn suggests that the dominant outflow may be a more collimated structure oriented approximately East-West, and similar to the coronal line region,  than the ionisation cone usually suggested.

\section{Conclusions}

In agreement with other work (Packham et al 2005), we find that the nucleus of the Circinus galaxy is compact at 10~$\umu$m  and heavily obscured.  Simple fits to the spectrum of the core reveal warm emitting grains suffering absorption by cool silicate dust with $\tau_{9.7\umu{\rm m}} \ge$2.2.  This provides a lower limit  to the true silicate absorption depth and suggests   A$_V  >$ 30 mag.  \\[2mm]
Weak emission extends to 2 arcsec east and west of the nucleus, with the east side considerably more obscured than the west side. This extended emission arises from  warm, relatively featureless dust grains in the coronal line emitting region which are probably  heated by trapped emission line photons. Prominent [SIV] line emission is detected in the extended emission west of the nucleus, and appears to be less strongly peaked than the higher ionization coronal lines mapped by Maiolino et al (2000) and Prieto et al (2004), suggesting ionisation stratification in this region. \\[2mm]  
It appears that the [SIV] emission is concentrated in the extended coronal line region to the east and west of the nucleus, whilst the [NeII] line is emitted from a more extended region and probably mostly  arises from circumnuclear HII regions.  In turn this indicates that these HII regions are fairly low ionisation objects, similar to those in many other galaxy nuclei. \\[2mm]
The extended emission on the east side of the nucleus is more highly obscured than that on the west side with an additional optical depth $\tau_{9.7\umu{\rm m}} \sim$1.6.  It seems that both the continuum  dust emission and the [SIV] line emission suffer similar extinction, and the underlying emission has a high degree of symmetry. We argue that this obscuring material must be local, consistent with a dusty structure inclined to our line of sight so that it lies between the earth and the coronal zone on the east side, but lies behind the coronal zone on the west side. This structure extends to at least 2 arcsec east of the nucleus, though the column falls by a factor of 2 over this distance.   We speculate that this structure may be a much larger scale version of the compact inclined molecular ring mapped by Greenhill et al ( 2003). \\[2mm]
The PAH band emission does not appear to suffer significant extinction from this material and so must arise outside the nuclear region, as expected if the carriers of the PAH bands are destroyed by the AGN photon flux (Aitken \& Roche 1985). Only a small fraction of the PAH emission is admitted by the T-ReCS slits, suggesting that the bulk of the PAH flux detected by ISO (Moorwood et al 1996)  arises in the extended circumnuclear star-forming regions.  

\section*{Acknowledgments}

This work is based on observations obtained at the Gemini Observatory, which is operated by the
Association of Universities for Research in Astronomy, Inc., under a cooperative agreement
with the NSF on behalf of the Gemini partnership: the National Science Foundation (United
States), the Particle Physics and Astronomy Research Council (United Kingdom), the
National Research Council (Canada), CONICYT (Chile), the Australian Research Council
(Australia), CNPq (Brazil) and CONICET (Argentina).  CP and JTR acknowledge support from 
NSF grant 0206617 and EP acknowledges support from NASA grant NNG05-GD63DG.

We thank the staff of Gemini-S, and especially Tom Hayward, for their assistance in collecting these data.   We are grateful to the referee for a careful critique of the manuscript, which led to several improvements in the presentation of these results.

\label{lastpage}

\end{document}